\documentclass{llncs}

\usepackage{amssymb}
\usepackage{amsmath}
\usepackage{textcomp}
\usepackage{array}
\usepackage[latin5]{inputenc}
\usepackage{lineno}
\usepackage{color}
\usepackage{comment}

\usepackage{graphicx}

\newcommand{\ket}[1]{|#1\rangle}
\newcommand{\braket}[2]{\langle #1|#2\rangle}
\newcommand{\cent}[0]{\mbox{\textcent}}
\newcommand{\dollar}[0]{\$}

\title{Log-space counter is useful for unary languages by help of a constant-size quantum register}

\author{Abuzer Yakary{\i}lmaz\inst{2}}

\institute{
University of Latvia, Faculty of Computing, Raina bulv. 19, R\={\i}ga, LV-1586, Latvia \\
\email{abuzer@lu.lv}
}

\authorrunning{A. Yakary{\i}lmaz} 

\begin{document}

\maketitle

\begin{abstract}
The minimum amount of resources to recognize a nonregular language is a fundamental research topic in theoretical computer science which has been examined for different kinds of resources and many different models. In this note, we focus on unary languages and space complexity on counters. Our model is two-way one-counter automaton with quantum and classical states (2QCCA), which is a two-way finite automaton with one-counter (2DCA) augmented with a fixed size quantum register or a two-way finite automaton with quantum and classical states (2QCFA) augmented with a classical counter. It is known that any 2DCA using a sublinear space on its counter can recognize only regular languages \cite{DG82B}. In this note, we show that bounded-error 2QCCAs can recognize a non-regular unary language by using logarithmic space on its counters for the members. Note that it is still an open problem whether bounded-error 2QCFA can recognize a non-regular unary language.
\\ ~~ \\
\textbf{Keywords:} \textit{automata theory, counter automata, quantum automata, unary languages, non-regular languages, pebble automata}
\end{abstract}

\section{Background}

We assume the reader familiar with the basics of the quantum computation (see \cite{NC00}). Some part of the background are taken from a previous related work \cite{Yak13B}, in which the same models are defined.

Throughout the paper, $ \Sigma $ not containing $ \cent $ and $ \dollar $ denotes the input alphabet and $ \tilde{\Sigma} = \Sigma \cup \{ \cent,\dollar \} $. For a given string $ w $, $ |w| $ is the length of $ w $ and $ w_{i} $ is the $ i^{th} $ symbol of $ w $, where $ 1 \leq i \leq |w| $. The string $ \cent w \dollar $ is represented by $ \tilde{w} $. Moreover, $ \Theta $ is the set of $ \{0,\pm\} $. 

The models defined in the paper have a two-way infinite read-only input tape whose squares are indexed by integers. Any given input string, say $ w \in \Sigma^{*} $, is placed on the tape as $ \tilde{w} $ between the squares indexed by 1 and $ | \tilde{w} | $. The tape has a single head, and it can stay in the same position ($ \downarrow $) or move to one square to the left ($ \leftarrow $) or to the right ($ \rightarrow $) in one step. It must always be guaranteed that the input head never leaves $ \tilde{w} $. A counter is an infinite storage having two status, i.e. \textit{zero} ($ 0 $) or \textit{nonzero} ($ \pm $), and updated by a value from $ \{-1,0,1\} $ in one step. 

A two-way deterministic one-counter automaton (2DCA) is a two-way deterministic finite automaton with a counter. Formally, a 2DCA $ \mathcal{D} $ is a 6-tuple 
\[
	\mathcal{D} = (S,\Sigma,\delta,s_{1},s_{a},s_{r}),
\]
where $ S $ is the set of states, $ s_1 \in S $ is the initial state, $ s_a \in S $ and $ s_r \in S $ ($ s_a \neq s_r $) are the accepting and rejecting states, respectively, and $ \delta $  is the transition function governing the behaviour of $ \mathcal{D} $ in each step, i.e. 
\[
	\delta: S \times \tilde{\Sigma} \times \Theta \rightarrow S \times \{ \leftarrow,\downarrow,\rightarrow \} \times \{-1,0,1\}. 
\]
Specifically, $ \delta(s,\sigma,\theta) \rightarrow (s',d_i,c) $ means that when $ \mathcal{D} $ is in state $ s \in S $, reads symbol $ \sigma \in \tilde{\Sigma} $, and the status of its counter is $ \theta \in \Theta $, then it updates its state to $ s' \in S $, the position of the input head with respect to $ d_i \in \{ \leftarrow,\downarrow,\rightarrow \}$, and the value of the counter by $ c \in \{-1,0,1\} $.

At the beginning of the computation, $ \mathcal{D} $ is in state $ s_1 $, the input head is placed on symbol $ \cent $, and the value of the counter is set to zero. A configuration of $ \mathcal{D} $ on a given input string is represented by a triple $ (s,i,v) $, where $ s $ is the state, $ i $ is the position of the input head, and $ v $ is the value of the counter. The computation is terminated and the input is accepted (rejected) by $ \mathcal{D} $ when it enters to $ s_a $ ($ s_r $). 

A two-way finite state automaton with quantum and classical states \cite{AW02} (2QCFA) is a two-way finite state automaton using a constant-size quantum register. Note that, a two-way probabilistic automaton (2PFA) is obtained if the quantum register is replaced by a random number generator. The main difference between a random number generator and a constant-size quantum register is that quantum register can keep some information by its (pure) quantum state as well.\footnote{It was shown that 2QCFAs are more powerful than their probabilistic counterparts in the bounded-error setting by Ambainis and Watrous \cite{AW02}. In fact, they can recognize many interesting languages in the bound error setting \cite{YS10A,YS10B}, currently not know to be recognized by 2PFAs.}

The definition of 2QCFA was introduced in \cite{AW02} where two kinds of quantum operators can be applied to quantum register: Unitary operators and orthogonal measurements. The computation is governed classically. A quantum operator is determined by the classical part and the automaton can follow a different classical transition for each measurement outcome. In \cite{YS11A}, a similar quantum model, two-way quantum finite automaton with classical head (2CQFA), was introduced where the computation is governed quantumly and superoperators (see Figure \ref{fig:superoperators}) are the main quantum operators. 2QCFAs and 2CQFAs can simulate each other exactly. Here, we follow the original definition of 2QCFA by allowing them to use superoperators. Such capability does not increase the computational power of 2QCFAs (see discussion at the end of Section 2 of \cite{AW02}). If only rational amplitudes are allowed to use, on the other hand, it is still unknown whether 2QCFAs using rational superoperators can be simulated by the original 2QCFAs using only rational unitary and measurement operators. For other kind of amplitudes such as algebraic, computable, unrestricted, etc. such simulation can be easily obtained. In this paper, we restrict ourselves with only rational superoperators. Note that any such 2QCFA algorithm can be implemented by an original 2QCFA using algebraic unitary operators and integer projective measurements.

\begin{figure}[!ht]
	\centering
	\footnotesize
	\fbox{
	\begin{minipage}{0.95\textwidth}
		The most general quantum operator is a superoperator,
		which generalizes stochastic and unitary operators and also includes measurements.
		Formally, a superoperator $ \mathcal{E} $ is composed by a finite number of operation elements,
		$ \mathcal{E} = \{ E_{1}, \ldots, E_{k} \} $, satisfying that
		\begin{equation}
		\label{eq:completeness}
			\sum_{i=1}^{k} E_{i}^{\dagger} E_{i} = I,
		\end{equation}
		where $ k \in \mathbb{Z}^{+} $ and the indices are the measurement outcomes.
		When a superoperator, say $\mathcal{E}$, is applied to 
		the quantum register in state $\ket{\psi} $, i.e. $ \mathcal{E}(\ket{\psi}) $,
		we obtain the measurement outcome $i$ with probability 
		$ p_{i} = \braket{\widetilde{\psi_{i}}}{\widetilde{\psi_{i}}} $,
		where $\ket{\widetilde{\psi_{i}}}$, \textit{the unconditional state vector}, 
		is calculated as $ \ket{\widetilde{\psi}_{i}} = E_{i} \ket{\psi} $ and $1 \leq i \leq k$.
		Note that using unconditional state vector simplifies calculations in many cases.
		If the outcome $i$ is observed ($p_{i} > 0 $), the new state of the system 
		is obtained by normalizing $ \ket{\widetilde{\psi}_{i}} $, 
		which is $ \ket{\psi_{i}} = \frac{\ket{\widetilde{\psi_{i}}}}{\sqrt{p_{i}}} $.
		Moreover, as a special operator, the quantum register can be initialized to a predefined quantum state.
		This initialize operator, which has only one outcome, is denoted $ \acute{\mathcal{E}} $.
	\end{minipage}
	}
	\caption{The details of superoperators \cite{Yak13C}}
	\label{fig:superoperators}
\end{figure}

A 2QCFA $ \mathcal{Q} $ is a 8 tuple 
\[
	(S,Q,\Sigma,\delta,s_1,s_a,s_r,q_1),
\] 
where, apart from a classical model, there are two different components: $ Q $ is the state set of quantum register and $ q_1 $ is its initial state. Moreover, $ \delta = (\delta_q,\delta_c) $, where $ \delta_q $ governs the quantum part and $ \delta_c $ governs the classical part. In each step, firstly, $ \delta_q $ determines a superoperator depending on the current classical state ($ s \in S $) and scanning symbol ($ \sigma \in \tilde{\Sigma} $), i.e. $ \mathcal{E}_{s,\sigma} $, and then it is applied to the quantum register and one outcome, say $ \tau $, is observed. Secondly, the classical part of $ \mathcal{Q} $ is updated depending on $ s $, $ \sigma $, and $ \tau $, which is formally represented as $ \delta_c(s,\sigma) \overset{\tau} \rightarrow (s',d_i) $, where $ s' \in S $ is the new classical state and $ d_i \in \{ \leftarrow,\downarrow,\rightarrow \} $ is the update of the position of input tape. Note that $ \delta_c $ must be defined for each possible $ \tau $. 

A two-way one-counter automaton with quantum and classical states (2QCCA) is a 2DCA augmented with a constant-size quantum register or a 2QCFA having a classical counter. The formal definition of a 2QCCA is exactly the same as a 2QCFA. So, the transition functions of a 2QCFA ($ \delta_q $ and $ \delta_c $) can be extended for a 2QCCA with the following modifications:
\begin{itemize}
	\item The superoperator is determined by also the status of the counter ($ \theta \in \Theta $), i.e. $ \mathcal{E}_{s,\sigma,\theta} $.
	\item The classical part of $ \mathcal{Q} $ is updated depending on $ s $, $ \sigma $, $ \theta $, and $ \tau $, which is formally represented as $ \delta_c(s,\sigma,\theta) \overset{\tau} \rightarrow (s',d_i,c) $, where $ s' \in S $ is the new classical state, $ d_i $ is the update of the position of input tape, and $ c \in \{-1,0,1\}  $ is the update on the counter.
\end{itemize}  

\section{Main result}

\newcommand{\Musquare}{\mathcal{USQUARE}}
\newcommand{\Msquare}{\mathcal{SQUARE}}
\newcommand{\Mupower}{\mathcal{UPOWER}}
\newcommand{\Mpower}{\mathcal{POWER}}

Recently, Yakary{\i}lmaz \cite{Yak13B} introduce a new programming technique for 2QCCAs (and also for 2QCFAs with a pebble) and it was shown that  
\[
	\mathtt{USQAURE} = \{ a^{n^2} \mid n \geq 1 \}
\]
can be recognized by a 2QCCA for any error bound by using a $ \sqrt{n} $-space on its counter for the members. The main idea is as follows. Let $ w = a^m $ be the input, where $ m \geq 1 $. Otherwise, it is rejected immediately. The 2QCCA, say $ \Musquare $, tries to mark the input from 1 to $ m $ iteratively by using its (classical) counter in a FOR-LOOP. That is, on the $ i^{th} $ iteration, the value of the counter is $ i $ on the left end-marker. Then $ \Musquare $ reads $ i $ $ a $'s by decrement the value of the counter by 1. It arrives on the $ i^{th} $ $ a $ when the counter becomes empty. After that, $ \Musquare $ firstly reads $ i $ $ a $'s again in reverse direction (by moving to the left end-marker) and then read the whole input, $ m $ $ a $'s, by moving to the right end-marker. In the first part, the counter value is set to $ i $ again, and, in the second part, the value of counter does not changed. In fact, $ \Musquare $ reads $ a^i \cent a^m $. We know that 2QCFAs can recognize 
\[
	\mathtt{SQAURE} = \{ a^n b^{n^2} \mid n \geq 1 \}
\]
for any error bound such that the members are accepted exactly and the non-members are rejected with high probability \cite{YS10B}. Let $ \Msquare $ be such 2QCFA rejecting the non-members with a probability at least $ \frac{2}{3} $. $ \Musquare $ executes $ \Msquare $ on $ a^i \cent a^m $ to test whether $ i^2 $ is equal to $ m $. $ \Msquare $ needs to read $ a^i \cent a^m $ many times and it can be provided by $ \Musquare $ easily since the value of $ i $ is stored on the counter. (Note that $ \Msquare $ reads its input from left to right in an infinite loop and the computation terminates with probability 1 in exponential expected time.) $ \Musquare $ passes to the next iteration only if $ \Msquare $ gives the decision of ``rejection''. So, if $ i^2 = m $, then $ \Musquare $ never passes to the $ (i+1)^{th} $ iteration since $ \Msquare $ never gives the decision of ``rejection''. $ \Musquare $ terminates the FOR-LOOP if $ \Msquare $ gives the decision of ``acceptance''. 

If $ w $ is a member of $ \mathtt{USQAURE} $, then $ \Musquare $ reaches the end-marker with a probability at least $ \left( \frac{2}{3} \right)^m $. At this point, $ \Musquare $ rejects $ w $. After termination of the FOR-LOOP, $ \Musquare $ accepts the input with probability $ \left( \frac{1}{3} \right)^{2m}  $. Thus,
\begin{itemize}
	\item any member of $ \mathtt{USQUARE} $ is accepted with a probability $ \left( \frac{1}{3} \right)^{2m}  $, and
	\item any non-member is rejected with a probability at least $ 4^m $ greater than the accepted probability.
\end{itemize}
So, by executing the above procedure in an infinite loop, $ \Musquare $ accepts any member of $ \mathtt{USQAURE} $ exactly, and rejects any non-member with a probability at least $ \frac{4}{5} $. By using conventional probability amplification techniques, the rejecting probability can be bounded below arbitrary close to 1. As can be easily verified, for the members, the counter value never exceeds $ \sqrt{m} = i $. For the non-members, on the other hand, $ \Musquare $ uses linear space.

Since the marking of the input described above can be implemented by a pebble as well, we can follow that $ \mathtt{USQAURE} $ can be recognized by a 2QCFA with a pebble for any error bound such that the pebble is moved only in $ \sqrt{|w|} $ times for any $ w \in \mathtt{USQUARE} $.

Similar to $ \mathtt{SQUARE} $,
\[
	\mathtt{POWER} = \{a^n b^{2^n} \mid n \geq 1 \}
\]
can also be recognized by 2QCFAs for any error bound \cite{YS10A,YS10B}. Some other languages recognized by bounded-error 2QCFAs listed in \cite{YS10B} are as follows:
\begin{itemize}
	\item $ \mathtt{TWIN} = \{ wcw \mid w \in \{a,b\}^* \} $
	\item $ \mathtt{MULT} = \{ x\#y\#z \mid x,y,z \in \{0,1\}^* \mbox{ and }x \times y  = z \} $ 
	\item All polynomial languages \cite{Tu82} defined as
	\[ 
		\{a_{1}^{n_{1}} \cdots a_{k}^{n_{k}} b_{1}^{p_{1}(n_{1},\ldots,n_{k})} \cdots b_{r}^{p_{r}(n_{1},\ldots,n_{k})} \mid p_{i}(n_{1},\ldots,n_{k}) \ge 0 \}, 
	\]
	where $ a_{1}, \ldots, a_{k},b_{1}, \ldots, b_{r} $ are distinct symbols, and each $ p_{i} $ is a polynomial with integer coefficients.
\end{itemize}
Actually, all these algorithms can be implemented by reading the input from left to right with a realtime head\footnote{The input head cannot stay on the same tape square more than a fixed number of steps.} in an infinite loop. We can call such models as restarting realtime finite automata with quantum and classical states (restarting rtQCFA). Knowledgeable readers can notice that restarting rtQCFA is a special case of rotating \cite{SS87,KKM12} 2QCFA, which is a special case of sweeping \cite{Sip80,KKM12} 2QCFA. Moreover, all these restarting rtQCFAs can be defined only with rational superoperators. We refer the reader also to \cite{Yak13C,Yak13A} for similar algorithms. Here, we will give a rational 2QCFA, say $ \Mpower $, (i.e., restarting rtQCFA) recognizing $ \mathtt{POWER} $ for any error-bound. After that, we describe a bounded-error 2QCCA for language
\[
	\mathtt{UPOWER} = \{ a^{2^n} \mid \geq 0 \}
\]
such that the 2QCCA uses logarithmic space on its counter for the members.

Let $ w \in \{a,b\}^* $ be the input. We can assume the input of the form $ a^mb^n $, where $ m,n>0 $. $ \Mpower $ rejects the input immediately, otherwise. The quantum register has three states: $ \ket{q_1}, \ket{q_2}, \ket{q_3} $. $ \Mpower $ encodes $ 2^m $ and $ n $ into amplitudes of $ \ket{q_2} $ and $ \ket{q_3} $, and then compare them by subtracting. If they are equal, then the resulting amplitude will be zero, and it is non-zero, otherwise. Based on this amplitude, the input is rejected. Since we will use only rational numbers, we can bound this rejecting probability from the below when it is non-zero. Note that it is zero only for the members. Thus, by creating a carefully tuned accepting probability, the members can be only accepted and the non-members can be rejected with a probability sufficient greater than the accepting probability. The technical details are given below.

The following procedure is executed in an infinite loop. In each iteration (round), the input is read from left to right in realtime mood. At the beginning of the round, the quantum state is set to $ \ket{\psi_0}=(1~~0~~0)^T $. In order to facilitate the calculations, the unconditional quantum state is traced as long as the current round is not terminated. When reading the left end-marker, $ \mathcal{E}_{\cent} = \{ E_{\cent,1},E_{\cent,2} \} $ is applied to the quantum register, i.e.
\[
	E_{\cent,1} = \dfrac{1}{2} \left( \begin{array}{rrr}
		1 & 0 & 0 \\
		1 & 0 & 0 \\
		0 & 0 & 2		
		\end{array}	 \right)
		\mbox{ and }
	E_{\cent,2} = \dfrac{1}{2} \left( \begin{array}{rrr}
		1 & 0 & 0 \\
		1 & 0 & 0 \\
		0 & 2 & 0		
		\end{array}	 \right),
\]
where (i) the current round continues if outcome ``1'' is observed, and (ii) the current round is terminated without any decision if outcome ``2'' is observed. Before reading $ a $'s, the quantum state is 
\[
	\ket{\widetilde{\psi_0}} = \dfrac{1}{2} \left( \begin{array}{c} 1 \\ 1 \\ 0 \end{array}	 \right).
\]
When reading an $ a $, $ \mathcal{E}_{a} = \{E_{a,1},E_{a,2} \} $ is applied to the quantum register, i.e.
\[
	E_{a,1} = \dfrac{1}{2} \left( \begin{array}{rrr}
		1 & 0 & 0 \\
		0 & 2 & 0 \\
		0 & 0 & 2		
		\end{array}	 \right)
		\mbox{ and }
	E_{a,2} = \dfrac{1}{2} \left( \begin{array}{rrr}
		1 & 0 & 0 \\
		1 & 0 & 0 \\
		1 & 0 & 0		
		\end{array}	 \right),
\]
where (i) the current round continues if outcome ``1'' is observed, and (ii) the current round is terminated without any decision if outcome ``2'' is observed. Before reading $ b $'s, the quantum state is 
\[
	\ket{\widetilde{\psi_m}} = \left( \dfrac{1}{2} \right) ^ {m+1} \left( \begin{array}{c} 1 \\ 2^m \\ 0 \end{array}	 \right).	
\]
When reading a $ b $, $ \mathcal{E}_{b} = \{E_{b,1},E_{b,2},E_{b,3} \} $ is applied to the quantum register, i.e.
\[
	E_{b,1} = \dfrac{1}{2} \left( \begin{array}{rrr}
		1 & 0 & 0 \\
		0 & 1 & 0 \\
		1 & 0 & 1		
		\end{array}	 \right),
	E_{b,2} = \dfrac{1}{2} \left( \begin{array}{rrr}
		1 & 0 & -1 \\
		1 & 0 & 0 \\
		0 & 1 & 1		
		\end{array}	 \right),
	\mbox{ and }
	E_{b,3} = \dfrac{1}{2} \left( \begin{array}{rrr}
		0 & 1 & -1 \\
		0 & 1 & 0 \\
		0 & 0 & 0		
		\end{array}	 \right),
\]
where (i) the current round continues if outcome ``1'' is observed, and (ii) the current round is terminated without any decision if outcome ``2'' or ``3'' is observed. Before reading the right end-marker, the quantum state is 
\[
	\ket{\widetilde{\psi_{|w|}}} = \left( \dfrac{1}{2} \right) ^ {m+n+1} \left( \begin{array}{c} 1 \\ 2^m \\ n \end{array}	 \right).	
\]
When reading the right end-marker, $ \mathcal{E}_{\dollar} = \{E_{\dollar,1},E_{\dollar,2},E_{\dollar,3},E_{\dollar,4} \} $ is applied to the quantum register, i.e.
\[
	E_{\dollar,1} = \dfrac{1}{2k} \left( \begin{array}{rrr}
		1 & 0 & 0 \\
		0 & 0 & 0 \\
		0 & 0 & 0		
		\end{array}	 \right),
	E_{\dollar,2} = \dfrac{1}{2k} \left( \begin{array}{rrr}
		0 & 0 & 0 \\
		0 & k & -k \\
		0 & k & -k		
		\end{array}	 \right),
	E_{\dollar,3} = \dfrac{1}{2k} \left( \begin{array}{rrr}
		k_1 & 0 & 0 \\
		k_2 & 0 & 0 \\
		k_3 & 0 & 0		
		\end{array}	 \right),
	\mbox{ and }
\]
\[
	E_{\dollar,4} = \dfrac{1}{2k} \left( \begin{array}{rrr}
		k_4 & 0 & 0 \\
		0 & k & k \\
		0 & k & k		
		\end{array}	 \right),
\]
where $ k $ can be any non-negative integer; $ k_1,k_2,k_3, \mbox{ and } k_4 $ are integers satisfying $ k_1^2+k_2^2+k_3^2+k_4^2 = 4k^2-1 $; and, the actions based on the measurement outcomes are as follows:
\begin{itemize}
	\item the input is accepted if outcome ``1'' is observed,
	\item the input is rejected if outcome ``2'' is observed, and,
	\item the current round is terminated without any decision, otherwise.
\end{itemize}
Note that, we define $ k_i $'s to have a well-defined quantum operator and, due to Lagrange's four-square theorem, we know that each natural number can be represented as the sum of four integer squares, where $ 1 \leq i \leq 4 $. The analysis of the algorithm is as follows. If outcome ``1'' is observed, then the quantum state is
\[
	\ket{\widetilde{\psi_{|w|+1}}} = \left( \dfrac{1}{2} \right) ^ {m+n} \dfrac{1}{k} \left( \begin{array}{c} 1 \\ 0 \\ 0 \end{array}	 \right).
\]
That is, in a single round, the input is always accepted with probability
\[
	\left( \dfrac{1}{4} \right) ^ {m+n} \dfrac{1}{k^2}.
\]
If outcome ``2'' is observed, then the quantum state is
\[
	\ket{\widetilde{\psi_{|w|+1}}} = \left( \dfrac{1}{2} \right) ^ {m+n} \dfrac{1}{k} \left( \begin{array}{c} 0 \\ k (2^m-n) \\ k (2^m-n) \end{array}	 \right).
\]
That is, in a single round, the input is rejected with a probability
\[
	\left( \dfrac{1}{4} \right) ^ {m+n} \dfrac{1}{k^2} 2 \left( k (2^m-n) \right)^2,
\]
which is 
\begin{itemize}
	\item zero for any member and
	\item at least $ 2k^2 $ times greater than the accepting probability for any non-member.
\end{itemize}
Thus, we can say that $ \Mpower $ accepts any member exactly and rejects any non-member with a probability at least $ \frac{2k^2}{2k^2 + 1} $. Thus, the error bound can be arbitrary close to zero by setting $ k $ with an appropriate value.

A 2QCCA for $ \mathtt{UPOWER} $, say $ \Mupower $, can be defined similar to $ \Musquare$ (the 2QCCA given above for $ \mathtt{USQUARE} $). The pseudo-code of $ \Mupower $ is given below. Let $ w = a^m $ be the input. 
\begin{equation*}
	\label{program:usquare}
	\mbox{
		\begin{minipage}{0.9\textwidth}
			\footnotesize
			FOR $ i = 1 $ TO $ m $ 
				\\ \hspace*{15pt}
					RUN $ \Mpower $ on $ w' = a^{i}b^{m} $
					\\ \hspace*{30pt}					
					IF $ \Mpower  $ \underline{accepts} $ w' $ THEN \textbf{TERMINATE} FOR-LOOP
					\\ \hspace*{30pt}
					IF $ \Mpower  $ \underline{rejects} $ w' $ AND $ i=m $ THEN \textbf{REJECT} the input
			\\
			END FOR
			\\
			\textbf{ACCEPT} $ w $ with a nonzero probability at most $ \left( \frac{1}{2k^2+1} \right)^{m} $
			\\
			\textbf{RESTART} the algorithm
		\end{minipage}			
	}
\end{equation*}
For the members of $ \mathtt{UPOWER} $, the decision of ``rejection'' is never given in FOR-LOOP. Therefore, they are accepted exactly. For the non-members, the input is rejected with a probability at least $ \left( \frac{2k^2}{2k^2+1} \right)^m $ at the end of a FOR-LOOP. Since the input can be accepted with a probability at most $ \left( \frac{1}{2k^2+1} \right)^{m} $ after a FOR-LOOP, the rejecting probability is at least $ k^{2m} $ times greater than the accepting probability after a FOR-LOOP. Therefore, any non-member is rejected with a probability at least $ \frac{2k^2}{2k^2+1} $. It is clear that for the members, the counter value never exceeds $ \log( |w| )$, so the space complexity is logarithmic for the members.

Once getting the details of the above algorithms, it is quite straightforward to show that each of the following languages can be recognized by 2QCCAs.
\begin{itemize}
	\item $ \mathtt{POLY(p)} = \{ a^{p(n)} \mid n \geq 1 \} $,
	\item $ \mathtt{POWER(m)} \{ a^{m^n} \mid n \geq 1 \} $, and
	\item $ \mathtt{POLY\mbox{-}POWER(p,m)} = \{ a^{ p(n)m^n} \mid n \geq 1  \} $,  
\end{itemize}
where $ p $ is a polynomial such that $ p(\mathbb{Z^+}) >0 $ and $ m > 2 $. Curious readers can also obtain their own combinations. As described before, the counter is used to mark the input iteratively in the above 2QCCA algorithms. Since such an iteration can be easily implemented by using a pebble, all languages above can be recognized by bounded-error 2QCFAs with a pebble. A counter can also iteratively mark the input as $ 1,n,n^2,n^3,\ldots $ for some integer $ n \geq 2 $. Thus, we can define new languages recognized by bounded-error 2QCCAs by replacing $ n $ with $ 2^n $ in the languages given above. For example, 
\[
	\left\{ a^{2^{2^n}} \mid n \geq 1 \right\}, ~~ 
	\left\{ a^{3^{5^n}} \mid n \geq 1 \right\}, ~~
	\left\{ a^{2^n3^{2^n}} \mid n \geq 1 \right\}, \mbox{ etc.}
\]
All such languages can be recognized by 2QCCAs for any error bound such that the members are accepted exactly and the non-members are rejected with high probability. Currently, we do not know any bound-error 2QCFAs with a pebble for such languages.

\section{Concluding remarks}

In \cite{DG82B}, it was shown that any unary language recognized by a two-way deterministic pushdown automaton using sublinear space on its stack is regular. Therefore, 2DCAs using sublinear space cannot recognize any nonregular unary languages. Currently, we do not known whether nondeterminism, alternation, or using random choices can help. 

For automata without counters, it is still open whether 2QCFAs can recognize a non-regular unary language. But, it is known that 2PFAs cannot recognize a non-regular unary language with bounded-error \cite{Kan91B}. Moreover, $ \mathtt{USQUARE} $ is a nonstochastic language \cite{Tur81}, not recognizable by 2PFAs with unbounded-error. Indeed, it is not known whether using a counter can help for 2PFAs. 

\section{Acknowledgements}
We thank Alexander Okhotin, Holger Petersen, and Klaus Reinhardt for their answers to our questions on the subject matter of this paper.

\bibliographystyle{alpha}
\bibliography{tcs}

\end{document}